\documentclass[10pt,a4paper]{article}
\usepackage[english]{babel}
\usepackage{amsmath}
\date{}
\title{ Anharmonic Noncommutative Oscillator at Finite Temperature }
\author{ H. Sarvari Karaj-Abad, A. Jahan\\Research Institute for Astronomy and Astrophysics of Maragha
(RIAAM)\\ Maragha, P. O. Box: 55134 - 441, IRAN\\ jahan@riaam.ac.ir}
\begin{document}
\maketitle
\begin{abstract}
Classical and quantum anharmonic noncommutative oscillators with a quartic self-interacting potential are considered and the effect of self-interaction term on the free energy and partition function of both models is calculated to first order in coupling constant.
\end{abstract}
\section{Introduction}
A $d$-dimensional ($2\leq d$) noncommutative oscillator is an oscillator with a set of mutually noncommutating spatial degrees of freedom. In recent years, the dynamics and thermodynamics of the classical and quantum noncommutative harmonic oscillators have been studied by several authors mainly due to the renewed interest in the noncommutative field theories [1-16]. The central subject in most of these studies is the harmonic oscillator and the effects due to the presence of anharmonic terms are neglected by most authors. The present letter is an attempt to consider the effect of anharmonic terms on the thermodynamic of a noncommutative harmonic oscillator. To this end we consider a two dimensional noncommutative harmonic oscillator perturbed by a quartic potential. The paper is organized as follow:\\
In next section, we derive the partition function and free energy of a classical model. It is found that the thermodynamic of a classical harmonic oscillator is not influenced by the noncommutativity of its coordinates. But in presence of the anharmonic terms the noncommutativity affects the thermodynamic quantities. In Sec. 2, the partition function and free energy of a quantum noncommutative anharmonic oscillator is calculated. At zero temperature, the free energy yields the ground-state energy of the model. The peculiar behavior of the quantum model emerges when one fails to get the analogous results for the usual (commutative) anharmonic oscillator starting from the noncommutative one.
\section{Classical Anharmonic Oscillator}
The model we assume for an anharmonic oscillator has the following Hamiltonian
\begin{eqnarray}\label{9}
\nonumber
H&=&H_0+H_{int}\\
&=&\frac{\textbf{p}^{2}}{2m}+\frac{1}{2}m\omega^{2}\textbf{x}^{2}+\frac{\lambda}{4!}\textbf{x}^{4}
\end{eqnarray}
with $\textbf{x}=(x_1,x_2)$ and $\textbf{p}=(p_1,p_2)$. The coupling constant is denoted by $\lambda$. The noncommutative version of this model is
\begin{equation}\label{9}
H=\frac{\textbf{p}^{2}}{2m}+\frac{1}{2}m\omega^{2}\widehat{\textbf{x}}^{2}+\frac{\lambda}{4!}\widehat{\textbf{x}}^{4}
\end{equation}
with generalized canonical structure given [1]
\begin{eqnarray}
\big\{\widehat x_i,\widehat x_j\big\}&=& \theta\epsilon_{ij}\quad\quad i,j=1,2 \\
\big\{\widehat x_i,p_j\big\}  &=& \delta_{ij} \\
\big\{p_i,p_j\big\}  &=& 0
\end{eqnarray}
Here $\theta$ stands for noncommutativity parameter. The antisymmetric tensor $\epsilon_{ij}$ normalized as $\epsilon_{12}=1$. The standard canonical structure can be maintained with the aid of the so-called Bopp shift, namely [1]
\begin{equation}\label{1}
\widehat x_i\rightarrow x_i= \widehat x_i+\frac{1}{2}\theta\sum_{j=1}^2\epsilon_{ij}p_j
\end{equation}
Thus due to the Bopp shift, equation (3) modifies to the standard form, i.e.
\begin{equation}\label{1}
\{\widehat x_i,\widehat x_j\}=\theta_{ij}\longrightarrow\{ x_i, x_j\}=0
\end{equation}
Using the Bopp shift, one arrives at
\begin{eqnarray}\label{1}
\widehat{\textbf{x}}^{2}&\longrightarrow&\textbf{x}^{2}=\frac{1}{4}\theta^{2}\textbf{p}^{2}+\textbf{x}^{2}-\theta L_{z}\\
\widehat{\textbf{x}}^{4}&\longrightarrow&\textbf{x}^{4}=\bigg(\frac{1}{4}\theta^{2}\textbf{p}^{2}+\textbf{x}^{2}-\theta L_{z}\bigg)^2
\end{eqnarray}
where $L_z=x_1p_2-x_2p_1$. Inserting the equations (8) and (9) in (2), give rise to the free and interaction Hamiltonians of the form
\begin{eqnarray}\label{5}
H_{0}^{\theta}&=&\frac{\kappa}{2m}\textbf{p}^{\,2}+\frac{1}{2}m\omega^{2}\textbf{x}^{\,2}
+\frac{1}{2}m\omega^{2}\vec{\theta}\cdot{}\textbf{x}\times\textbf{p}\\
H_{int}^{\theta}&=&\frac{\lambda}{4!}\bigg[\textbf{x}^4-2\theta\textbf{x}^{\,2} L_z+\theta^2L^2_z
+\frac{1}{2}\theta^{2}\textbf{p}^{2}\textbf{x}^2-\frac{1}{2}\theta^3\textbf{p}^{2}L_z+\frac{1}{16}\theta^{4}\textbf{p}^4\bigg]
\end{eqnarray}
with the total Hamiltonian given by
\begin{equation}\label{1}
H^\theta=H^\theta_0+H^\theta_{int}
\end{equation}
In (10) we have introduced $\kappa=1+\frac{1}{4}m^2\omega^2\theta^2$ and $\vec\theta=(0,0,\theta)$. It is clear that equation (12) reproduces the Hamiltonian (1) when $\theta\rightarrow 0$. In equations (10), (11) and (12), the subscript $\theta$ appears to indicate the dependence of the results on the noncommutativity parameter. To avoid the notational complexity, we shall discard it in the following calculations.
\subsection{Partition function}
The classical partition function is
\begin{eqnarray}\label{1}
\nonumber
Z(\beta)&=&\textrm{Tr}e^{-\beta( H_0+H_{int})}\\
&=&\int_{-\infty}^{+\infty}{d^2{p}}\int_{-\infty}^{+\infty}{d^2{x}} \,e^{-\beta( H_0+H_{int})}
\end{eqnarray}
To evaluate this integral, we introduce the generating function defined to be
\begin{eqnarray}
\nonumber
G[\textbf{J}_{x},\textbf{J}_{p}]&=&\int_{-\infty}^{+\infty}{d^2{p}}\int_{-\infty}^{+\infty}{d^2{x}}e^{-\beta({H_{0}}+\scriptsize{\textbf{J}_{x}}\cdot\textbf{x}
+\textbf{J}_{p}\cdot\textbf{p})}\\
&=&\Big(\frac{2\pi}{\beta\omega}\Big)^2 e^{\frac{\beta}{2m}[\frac{\kappa}{\omega{2}}\scriptsize{\textbf{J}}_{x}^{2}+m^{2}\textbf{J}_{p}^{2}-m^{2}\vec\theta\cdot(\scriptsize{\textbf{J}}_{x}\times\textbf{J}_{p})]}
\end{eqnarray}
which could be evaluated straightforwardly with the help of Gaussian integral
\begin{equation}\label{1}
\int_{-\infty}^{+\infty}{d{x}}e^{-ax^2+bx}=e^{\frac{b^2}{4a}}\sqrt{\frac{\pi}{a}}
\end{equation}
Then equation (14) allows us to write
\begin{equation}\label{6}
\textrm{Tr}\Big(x_{i}^{n}p_{j}^{m}e^{-\beta H_{0}}\Big)=\frac{1}{\beta^{n+m}}\frac{\partial^{n+m}}{\partial^n{J}_{x,i}\partial^{m}{J}_{p,j}}G[\textbf{0},\textbf{0}]
\end{equation}
The partition function can be calculated perturbatively  starting from
\begin{eqnarray}
\nonumber
Z&=&\textrm{Tr}e^{-\beta( H_0+H_{int})}\\\nonumber
&=&\textrm{Tr}e^{-\beta H_0}-\beta\textrm{Tr}\Big(e^{-\beta H_0}H_{int}\Big)+O(\lambda^2)\\
&=&Z_0+Z_1
\end{eqnarray}
When the anharmonic term is absent ($\lambda=0$), the partition function becomes [1]
\begin{eqnarray}\label{1}
\nonumber
Z_0&=&\int_{-\infty}^{+\infty}{d^2{p}}\int_{-\infty}^{+\infty}{d^2{x}} \,e^{-\beta( \frac{\kappa}{2m}\textbf{p}^{\,2}+\frac{1}{2}m\omega^{2}\textbf{x}^{\,2}
+\frac{1}{2}m\omega^{2}\vec{\theta}\cdot{}\textbf{x}\times\textbf{p})}\\
&=&\Big(\frac{2\pi}{\beta\omega}\Big)^{2}
\end{eqnarray}
The integral is straightforwardly calculated using (15). Alternatively, from (14) one finds
\begin{eqnarray}\label{1}
\nonumber
Z_0&=&G[\textbf{0},\textbf{0}]\\
&=&\Big(\frac{2\pi}{\beta\omega}\Big)^{2}
\end{eqnarray}
We observe that the result is independent of the noncommutativity parameter $\theta$. So the partition function of a classical noncommutative harmonic oscillator is unaffected by the noncommutativity [1]. To obtain the the first order partition function $Z_1$, from (11), (16) and (17) we find
\begin{eqnarray}\label{1}
\textrm{Tr}\Big({x}^{4}_ie^{-\beta H_{0}}\Big)&=&\frac{3\kappa^{2}}{m^{2}\beta^2\omega^{4}}Z_0\\
\textrm{Tr}\Big({p}^{4}_ie^{-\beta H_{0}}\Big)&=&\frac{3m^2}{\beta^{2}}Z_0\\
\textrm{Tr}\Big(x_{i}^{2}x_{j}^{2}e^{-\beta H_{0}}\Big)&=&\frac{\kappa^2}{m^2\beta^{2}\omega^4}Z_0,\quad\quad (i\neq j)\\
\textrm{Tr}\Big(p_{i}^{2}p_{j}^{2}e^{-\beta H_{0}}\Big)&=&\frac{m^2}{\beta^2}Z_0,\;\quad\quad\quad\quad (i\neq j)
\end{eqnarray}
and
\begin{eqnarray}\label{7}
\textrm{Tr}\Big(x_{i}^{2}p_{j}^{2}e^{-\beta H_{0}}\Big)&=&\frac{1}{\beta^{2}\omega^{2}}\big[3\kappa-2-2(\kappa-1)\delta_{ij}\big]Z_0\\
\textrm{Tr}\Big(x_jp_jx_{i}p_{i}e^{-\beta H_{0}}\Big)&=&\frac{1}{\beta^{2}\omega^{2}}\big[1-\kappa+(2\kappa-1)\delta_{ij}\big]Z_0\\
\textrm{Tr}\Big(p_kx_jx_{i}^{2}e^{-\beta H_{0}}\Big)&=&-\frac{\kappa\theta}{2\beta^{2}\omega^{2}}\big(2\epsilon_{ik}\delta_{ij}+\epsilon_{jk}\big)Z_0\\
\textrm{Tr}\Big(x_kp_jp_{i}^{2}e^{-\beta H_{0}}\Big)&=&\frac{\theta m^2}{2\beta^{2}}\big(2\epsilon_{ik}\delta_{ij}+\epsilon_{jk}\big)Z_0
\end{eqnarray}
The terms of interaction  Hamiltonian (11) read now
\begin{eqnarray}\label{1}
\nonumber
\textrm{Tr}\Big( \textbf{x}^4e^{-\beta H_{0}}\Big)&=&\sum_{i}\textrm{Tr}\Big(x^4_ie^{-\beta H_{0}}\Big)+\sum_{i\neq j}\Big(x^2_ix^2_je^{-\beta H_{0}}\Big)\\
&=&\frac{8\kappa^2}{m^2\beta^2\omega^4}Z_0\\\nonumber
\textrm{Tr}\Big( \textbf{p}^4e^{-\beta H_{0}}\Big)&=&\sum_{i}\textrm{Tr}\Big(p^4_ie^{-\beta H_{0}}\Big)+\sum_{i\neq j}\Big(p^2_ip^2_je^{-\beta H_{0}}\Big)\\
&=&\frac{8m^2}{\beta^2}Z_0\\\nonumber
\textrm{Tr}\Big( L_z^2e^{-\beta H_{0}}\Big)&=&\sum_{ijmn}\epsilon_{ij}\epsilon_{mn}\textrm{Tr}\Big(x_ip_jx_mp_ne^{-\beta H_{0}}\Big)\\
&=&\frac{2(4\kappa-3)}{\beta^2\omega^2}Z_0\\\nonumber
\textrm{Tr}\Big(\textbf{x}^{\,2} L_ze^{-\beta H_{0}}\Big)&=&\sum_{ijk}\epsilon_{jk}\textrm{Tr}\Big(p_kx_jx^2_ie^{-\beta H_{0}}\Big)\\
&=&-\frac{4\kappa\theta}{\beta^2\omega^2}Z_0\\\nonumber
\textrm{Tr}\Big(\textbf{p}^{\,2} L_ze^{-\beta H_{0}}\Big)&=&\sum_{ijk}\epsilon_{kj}\textrm{Tr}\Big(x_kp_jp^2_ie^{-\beta H_{0}}\Big)\\
&=&-\frac{4\theta m^2}{\beta^2}Z_0\\\nonumber
\textrm{Tr}\Big( \textbf{x}^2\textbf{p}^2e^{-\beta H_{0}}\Big)&=&\sum_{ij}\textrm{Tr}\Big(x^2_ip^2_je^{-\beta H_{0}}\Big)\\
&=&\frac{4(2\kappa-1)}{\beta^2\omega^2}Z_0
\end{eqnarray}
where in (30) we have used $\epsilon_{ij}\epsilon_{kl}=\delta_{ik}\delta_{jl}-\delta_{il}\delta_{jk}$. Therefore, form (11) and (17) one is left with the first order partition function
\begin{eqnarray}\label{1}
\nonumber
Z_1(\beta)&=&-\beta\textrm{Tr}\Big(e^{-\beta H_0}H_{int}\Big)\\
&=&-\frac{1}{12}\frac{\lambda\theta^{2}}{\beta\omega^{2}}\frac{16\kappa^{2}-24\kappa+9}{\kappa-1}Z_0
\end{eqnarray}
When the noncommutativity vanishes, we get the correct limit, namely the partition function of the commutative oscillator
\begin{equation}\label{1}
Z_1(\beta)=-\frac{1}{3}\frac{\lambda}{\beta m^2 \omega^{4}}Z_0
\end{equation}
\subsection{Free Energy}
The Helmholtz free energy $F(\beta)$ is related to the (classical or quantum) partition function $Z(\beta)$ via
\begin{equation}\label{16}
e^{-\beta F}=Z(\beta)
\end{equation}
By expanding it as $F=F_0+F_1+...$, one obtains the zeroth and first order free energies as
\begin{eqnarray}\label{16}
F_0&=&-\frac{1}{\beta}\ln Z_0(\beta)\\\nonumber
F_1&=&\frac{\textrm{Tr}\Big(e^{-\beta H_0}H_{int}\Big)}{\textrm{Tr}e^{-\beta H_0}}\\
&=&-\frac{1}{\beta}\frac{Z_1}{Z_0}
\end{eqnarray}
Therefore, for the first order free energy with the partition functions (18) and (34) one obtains
\begin{eqnarray}\label{19}
F_1(\beta)=\frac{1}{12}\frac{\lambda\theta^{2}}{\beta^2\omega^{2}}\frac{16\kappa^{2}-24\kappa+9}{\kappa-1}
\end{eqnarray}
Once again, when the noncommutativity disappears, the first order free energy of a usual anharmonic oscillator is recovered
\begin{eqnarray}\label{19}
\lim_{\theta\rightarrow 0}F_1=\frac{1}{3}\frac{\lambda}{\beta^2 m^2 \omega^{4}}
\end{eqnarray}
\section{Quantum Anharmonic Oscillator}
The noncommutative quantum oscillator is characterized by a generalized canonical structure of the form
\begin{eqnarray}
\big[\widehat x_i,\widehat x_j\big]&=& i\theta_{ij}\quad\quad(\hbar=1)\\\
\big[\widehat x_i,p_j  \big]&=& i\delta_{ij} \\\
\big[p_i,p_j\big]&=& 0
\end{eqnarray}
Once again, the Bopp shift of the form [1, 2]
\begin{equation}\label{1}
\widehat x_i\rightarrow x_i= \widehat x_i+\frac{i}{2}\theta\sum_{j=1}^2\epsilon_{ij}p_j
\end{equation}
allows us to shift from the noncommutative coordinates to commutative degrees of freedom fulfilling $\big[ x_i,x_j\big]=0$. Now to consider the quantum noncommutative anharmonic oscillator, we follow [3] to introduce the quantum mechanical operator $K$ as
\begin{equation}\label{1}
K=\frac{\theta^{2}}{4}\textbf{p}^{2}+\textbf{x}^{2}-\theta L_{z}
\end{equation}
Thus upon implementing the Bopp shift, Eq. (44), and using (9) one obtains
\begin{equation}\label{1}
\widehat{\textbf{x}}^{4}\longrightarrow\textbf{x}^{4}=K^2
\end{equation}
So, the Hamiltonian (2) for a quantum anharmonic oscillator in terms of operator $K$ becomes
\begin{equation}\label{11}
H=H_{0}+\frac{\lambda}{4!}K^{2}
\end{equation}
which implies $H_{int}=\frac{\lambda}{4!}K^{2}$. For the eigen-ket $|n_{+},n_{-}\rangle$ the operator $K$ and Hamiltonian $H_0$ satisfy [3-5]
\begin{eqnarray}\label{1}
K|n_{+},n_{-}\rangle&=&\theta(2n_{-}+1)|n_{+},n_{-}\rangle\\
H_0|n_{+},n_{-}\rangle&=&\omega\big[(\sqrt{\kappa}+\sqrt{\kappa-1}\,)n_{+}+(\sqrt{\kappa}-\sqrt{\kappa-1}\,)n_{-}
+\sqrt{\kappa}\,\big]|n_{+},n_{-}\rangle
\end{eqnarray}
The eigen-ket $|n_{+},n_{-}\rangle$ is given by
\begin{equation}\label{1}
|n_{+},n_{-}\rangle=\frac{(a^{\dagger }_+)^{n_+}(a^{\dagger }_-)^{n_-}}{\sqrt{n_+!n_-!}}|0,0\rangle
\end{equation}
where $|0,0\rangle$ is the vacuum state of a two dimensional harmonic oscillator and $a_{\pm}^\dagger$ stands for the creation operator. Therefore, equations (48) and (49) imply that the Hamiltonian (47) is diagonal with respect to the state $|n_{+},n_{-}\rangle$. But this not always the case and as shown in [3, 4] in the absence of the harmonic term in (1), i.e. $\omega=0$ , there is a non-diagonal term given by $-\frac{2}{m\theta^2}\textbf{x}^2$.
From (48) the first order shift in the ground-state energy is found to be
\begin{eqnarray}\label{1}
\nonumber
\Delta E_{0,0}&=&
\langle 0,0|H_{int}|0,0\rangle\\
&=&\frac{\lambda}{4!}\theta^2
\end{eqnarray}
When the noncommutativity disappears the above result vanishes. Thus it does not lead to the shift in the ground-state energy of the commutative model. This peculiar behavior of equation (51) is a generic feature of the noncommutative models [4].
\subsection{Partition Function}
On using (49), the free part of the Hamiltonian yields the zeroth-order order quantum mechanical partition function [1-3]
\begin{eqnarray}\label{13}
\nonumber
Z_0&=&\textrm{Tr}e^{-\beta H_0}\\\nonumber
&=&\sum_{n_{+},n_{-}=0}^{\infty}\langle n_{+},n_{-}|e^{-\beta H_{0}}|n_{+},n_{-}\rangle\\
&=&\frac{1}{4\sinh\frac{\alpha_+}{2}
\sinh\frac{\alpha_-}{2}}
\end{eqnarray}
where we have defined $\alpha_{\pm}={\beta\omega}\big(\sqrt{\kappa}\pm\sqrt{\kappa-1}\big)$. With $H_{int}=\frac{\lambda}{4!}K^2$ from (17) we find the first order partition function as
\begin{eqnarray}\label{13}
\nonumber
Z_1&=&-\frac{1}{4!}\beta\lambda\sum_{n_{+},n_{-}=0}^{\infty}\langle n_{+},n_{-}|e^{-\beta H_{0}}K^2|n_{+},n_{-}\rangle\\
&=&-\frac{1}{4!}\lambda\beta\theta^{2}
\bigg(1+\frac{2}{\sinh^2\frac{\alpha_-}{2}}\bigg)Z_0
\end{eqnarray}
where we have used (47) and
\begin{equation}\label{1}
\textrm{Tr}\Big(e^{-\beta H_0}n_-^pn_+^q\Big)=(-1)^{p+q}e^{-\frac{\alpha_-+\alpha_+}{2}}\frac{\partial^{p+q}}{\partial\alpha_-^p\partial\alpha_+^q}\frac{1}{1-e^{-\alpha_-}}\frac{1}{1-e^{-\alpha_+}}
\end{equation}
\subsection{Free Energy}
From (52), the zeroth-order free energy reads
\begin{eqnarray}
F_0
&=&\frac{1}{\beta}\ln \Big(4\sinh\frac{\alpha_+}{2}
\sinh\frac{\alpha_-}{2}\Big)
\end{eqnarray}
At zero temperature, the above expression reproduce the unperturbed ground-state energy [1, 2]
\begin{equation}\label{1}
\lim_{\beta\rightarrow\infty}F_0(\beta)=\omega\sqrt\kappa
\end{equation}
The first-order free energy can be calculated straightforwardly with the aid of (38) and (53). We arrive at
\begin{eqnarray}\label{20}
F_1&=&\frac{1}{4!}\lambda\theta^{2}
\bigg(1+\frac{2}{\sinh^2\frac{\alpha_-}{2}}\bigg)
\end{eqnarray}
Again when ${\beta\rightarrow\infty}$, the above expression leads to the first order shift in the ground-state energy, in agreement with (51)
\begin{equation}\label{1}
\lim_{\beta\rightarrow\infty}F_1(\beta)=\frac{\lambda}{4!}\theta^2
\end{equation}

\section{Conclusion}
We considered a two dimensional noncommutative harmonic oscillator perturbed by a quartic potential. Then we derived the partition function and free energy of the classical model up to first order in coupling constant. The zeroth order partition function does not depend on the noncommutativity and coincides with the partition function of the usual commutative harmonic oscillator. The effect of the noncommutativity appears when one considers the first (or higher) order partition function. In contrast to the classical model, the zeroth order partition function of a quantum oscillator is affected by the noncommutativity. We obtained the first order free energy of the quantum oscillator. At zero temperature, it leads to the first order shift in the ground state energy of the model.


\begin{thebibliography}{99}
\bibitem{1} I. Jabbari, A. Jahan and Z. Rizai, Turk. J. Phys. \textbf{33}, 149 (2009) [Updated version at: http://arxiv.org/abs/1201.0827].
\bibitem{1} A. Jahan, Braz. J. Phys. \textbf{38}, 144 (2008).
\bibitem{1}  J. M. Carmona, J. L. Cortes, J. Gamboa and F. Mendez,   Int. J. Mod. Phys. \textbf{A 17}, 2555 (2002).
\bibitem{1} J. Gamboa, M. Loewe and J. Rojas, Phys. Rev. \textbf{D 64} 067901, (2001).
\bibitem{1} J. M. Carmona, J. L. Cortes, J. Gamboa and F. Mendez,  Mod. Phys. Lett. \textbf{A 16}, 2075 (2001).
\bibitem{1} B. Dragovic and Z. Rakic, Theor. Math. Phys. \textbf{140}, 1299 (2004).
\bibitem{1} B. Dragovic and Z. Rakic, Path Integral Approach to Noncommutative Quantum Mechanics, arXiv: hep-th/0401198
\bibitem{1} H. Benzaira, M. Meradc, T. Boudjedaad and A. Makhlouf, Z. Naturforsch. \textbf{67 A}, 77 (2012).
\bibitem{1} L. Mai-Lin, Z Ya-Bin, Y Rui-Lin and Z Fu-Lin, Chinese. Phys. \textbf{C 37}, 063106 (2013).
\bibitem{1} A. Smailagic and E. Spallucci, J.Phys. \textbf{A 35} L363 (2002).
\bibitem{1} E. Abreu, M. Marcial, A. Mendes and W. Oliveira, JHEP \textbf{11}, 138 (2013).
\bibitem{1} J. Geloun, S. Gangopadhyay and F. Scholtz, Europhys. Lett. \textbf{86}, 51001 (2009).
\bibitem{1} P. Giri and P. Roy, Eur. Phys. J. \textbf{C 57}, 835 (2008).
\bibitem{1} A. Hatzinikitas and I. Smyrnakis, J. Math. Phys. \textbf{43}, 113 (2002).
\bibitem{1} I. Dadic, L. Jonke and S. Meljanac, Acta Phys. Slov. \textbf{55}, 149 (2005).
\bibitem{1} S. A. Alavi, Prob. Atomic Sci. Technol. \textbf{3}, 301 (2007).

\end{thebibliography}
\end{document}